\documentclass[a4paper, 10pt, conference]{ieeeconf}%
\usepackage{graphics}
\usepackage{epsfig}
\usepackage{mathptmx}
\usepackage{times}
\usepackage{amsmath}
\usepackage{amssymb}
\usepackage{graphicx}
\usepackage{amssymb}
\usepackage{epstopdf}
\usepackage{subfig}
\usepackage{amsfonts}
\providecommand{\U}[1]{\protect\rule{.1in}{.1in}}
\IEEEoverridecommandlockouts
\begin{document}

\title{{\LARGE \textbf{Low Speed Control of an Autonomous Vehicle Using a Hybrid
Fractional Order Controller}}}
\author{S. Hassan HosseinNia, In\'{e}s Tejado, Blas M. Vinagre, Vicente Milan\'{e}s,
Jorge Villagr\'{a} \thanks{This work has been partially supported by research
grants TRA2008-06602-C01 and TRA2008-06602-C02 (Spanish Ministry of Science
and Innovation).}\thanks{S. Hassan HosseinNia, In\'{e}s Tejado and Blas M.
Vinagre are with Dept. of Electrical, Electronic and Automation Engineering,
Industrial Engineering School, University of Extremadura, Badajoz, Spain.
{\small (email: \{hoseinnia,itejbal,bvinagre\}@unex.es)}}\thanks{Vicente
Milan\'{e}s, Jorge Villagr\'{a} are with the AUTOPIA Program at the Centre for
Automation and Robotics (CAR, UPM-CSIC), Ctra. Campo Real km 0.22, La
Poveda-Arganda del Rey, 28500 Madrid, Spain.
{\small {(e-mail:\{vicente.milanes,jorge.villagra\}@csic.es})}}}
\maketitle

\begin{abstract}
Highly non-linear vehicle dynamics plays an important role in autonomous driving systems, especially in congested traffic situations at very low speeds. Due to this fact, accurate controllers are needed in order to ensure safety during navigation. In this paper, based on a previous fractional order speed control, an improved fractional order control is presented to control a commercial Citro\"{e}n C3 prototype --which has automatic driving capabilities-- at low speeds, which considers a hybrid model of the vehicle. Specifically, two different fractional order PI$^{\alpha}$ controllers are designed to act over the throttle and brake pedals, respectively. Concerning to the uncertain dynamics of the system during the brake action parameters are tuned to design a robust controller. In addition, the system is modeled as hybrid fractional order differential inclusions. Experimental and simulation results, obtained in a real circuit, are given to demonstrate the effectiveness of the proposed strategies for cruise control at low speeds.

\end{abstract}

\section{Introduction}

Research on traffic safety is continuously developing and carried out around
the world. In particular, the aim is to develop active systems, called
advanced driver assistance systems (ADAS), which will be able to prevent
accidents (see \cite{Vicente2011}). One of the systems included in commercial
vehicles to increase safety in carrying out driving-related tasks is cruise
control (CC). Standard cruise control (CC) will automatically adjust the
vehicle speed regarding to the desired reference velocity. Automotive sector
has included in its commercial vehicles some advances for driving in urban
areas. From the efficiency point of view, the Start$\&$Stop system in
\cite{pla2009} permits switching off the vehicle's engine when it is stopped
because of traffic lights or jams. Therefore, $CO_{2}$ emissions are
significantly reduced. Nevertheless, from the safety point of view, autonomous
systems capable of aiding the driver in case of congested traffic situations
still remain as an unsolved problem. The main difficulty arises from the
highly non-linear dynamics of vehicles at very low speeds.

Fractional order control (FOC), that is, the generalization to non-integer
orders of traditional controllers or control schemes, and its applications are
becoming an important issue since it translates into more tuning parameters
or, in other words, more adjustable time and frequency responses of the
control system, allowing the fulfillment of robust performances. It has been
applied, in a satisfactory way, in several automatic control applications (see
\cite{Monje10}, \cite{ChenFOC09} and references therein) leading to the
conclusion that FOC is preferred to other techniques (the better performance
of this type of controllers, in comparison with the classical PID ones, has
been demonstrated e.g. in \cite{Oustaloup91} and \cite{Podlubny99}). However,
FOC has not been applied to low speed control of autonomous vehicles,
expecting robust results.

With the above motivation, this paper deals with the design and implementation
of the CC of the commercial Citro\"{e}n C3 vehicle is addressed. As a matter
of fact, based on a hybrid model of the vehicle, which considers different
models for vehicle dynamics when accelerating and braking, two fractional
order PI$^{\alpha}$ controllers are designed for CC man\oe uvres at low
speeds. Moreover, the system will be modeled by fractional order hybrid
differential inclusions.

The rest of the paper is organized as follows. Section \ref{description}
briefly describes the modifications performed in the vehicle to act
autonomously on the throttle and brake pedals, as well as the dynamic
longitudinal model obtained for this kind of man\oe uvres. Section
\ref{CC}\ addresses the fractional order CC of the vehicle acting over the
throttle and brake pedals. Simulation and experimental results are given in
Section \ref{results} to validate the proposed CC. Finally, concluding remarks
are included in Section \ref{conclu}.

\section{Automatic Vehicle\label{description}}

To design the cruise control man\oe uvres at very low speeds, a
model of the automatic vehicle --a commercial convertible Citro\"{e}n C3
Pluriel (see Fig. \ref{Car_Pic})-- was obtained experimentally, which includes its dynamics when
accelerating and braking at very low speeds. This section briefly describes
the modifications performed in the vehicle to act autonomously on the throttle
and brake pedals, as well as its dynamic longitudinal model.
{\normalsize \begin{figure}[ptbh]
\begin{center}
{\normalsize \includegraphics[width=0.4\textwidth]{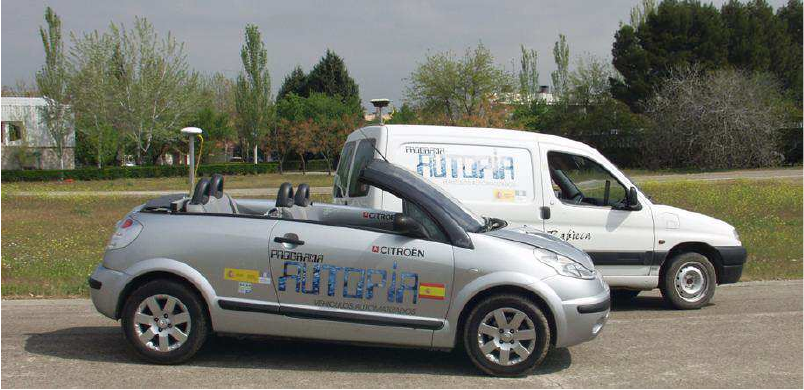} }
\end{center}
\caption{Experimental vehicle}%
\label{Car_Pic}%
\end{figure}}

\subsection{Description}

The vehicle control system for automatic driving follows the classical
perception-reasoning-action paradigm \cite{Onieva10},\cite{Vicente_2010a}. The
first stage is in charge of localizing as precisely and robustly as possible
the vehicle. To that end, the following subsystems are embedded in the vehicle

\begin{itemize}
\item A double-frequency global positioning system (GPS) receiver running in
real-time kinematic (RTK) carrier phase differential mode that supplies $2$cm
of resolution positioning at a refresh rate of $5$Hz.
\item A wireless local area network (IEEE $802.11$) support, which allows the
GPS to receive both positioning error corrections from the GPS base station
and vehicle and positioning information from the preceding vehicle.
\item An inertial measurement unit (IMU) Crossbow IMU 300CC placed close to
the centre of the vehicle to provide positioning information during GPS outages.
\item Car odometry supplied by a set of built-in sensors in the wheels, whose
measurements can be read by accessing the controller area network (CAN) bus of
the vehicle. This is implemented by means of a CAN Card $2.6$.
\end{itemize}

Thereafter, an on-board computer is in charge of requesting values from each
of the on-board sensors with which to compute the controller's input values. Finally, the devices that make possible to act on the throttle and brake of
the car are an electrohydraulic system capable of injecting pressure into the
car's anti-block braking system (ABS), and an analogue card which can send a
signal to the car's internal engine computer to demand acceleration or
deceleration. The electro-hydraulic braking system is mounted in parallel with
the original one. Two shuttle valves are installed connected to the input of
the anti-lock braking system (ABS) in order to keep the two circuits
independent. A pressure limiter tube set at $120$bars is installed in the
system to avoid damage to the circuits. Two more valves are installed to
control the system: a voltage-controlled electro-proportional pilot to
regulate the applied pressure, and a spool directional valve to control the
activation of the electrohydraulic system by means of a digital signal. These
two valves are controlled via an I/O digital-analogue CAN card. The voltage
for the applied pressure is limited to $4$V (greater values correspond to hard
braking and are not considered).

\subsection{Dynamic longitudinal model}

Due to the impossibility of obtaining the exact dynamics that describes the
vehicle, in this work the idea is to obtain a simple linear model of the
vehicle for the circuit wherein the experimental man\oe uvres will be performed.

The vehicle longitudinal dynamics can be simplified by a first order transfer
function \cite{Ines_2011} that relies the vehicle velocity and a proportional
voltage to the throttle angle:%
\begin{equation}
G(s)\simeq\frac{K}{s+p}=\frac{4.39}{s+0.1746}, \label{throttle}%
\end{equation}

Simple linear longitudinal models have been also used in \cite{Rodriguez07}
and \cite{Kamga96}. The reason why there is no need to use a more complex
model arises from the kind of man\oe uvres we perform in this work, as will be
stated from the experimental results.

Besides, vehicle dynamics in braking maneuvers can be given by an uncertain
first order transfer function that depends on the voltage applied to the brake
pedal \cite{Vicente_2010b}.
\begin{equation}
G(s)\simeq\frac{1}{\tau s+1},\label{brake_eq}%
\end{equation}
where the time constant $\tau$ varies with the action over the brake in the
interval $\tau\in\lbrack1.6,3.1]$s.

\section{Cruise Control\label{CC}}

This section presents a hybrid CC of the vehicle at low speeds based on the
different vehicle's dynamics when accelerating and braking. In particular, the
fractional order PI$^{\alpha}$ controller designed in \cite{Ines_2011} will be
used for the throttle action --it was designed to control the throttle and
brake pedals, but neglecting the dynamics during braking--, whereas the brake
will be controlled by a robust fractional order PI due to the system
uncertainty described previously. The motivation of improving that design by
considering a hybrid model of the vehicle mainly arises from its application
to ACC man\oe uvres, in which the adequate control of the brake pedal plays a
key role for the success of the whole test. Some considerations on the
switching of the controllers are also included.

The most important mechanical and practical requirement of the vehicle to take
into account during the design process is to obtain a smooth vehicle's
response so as to guarantee its acceleration to be less than the well-known
comfort acceleration, i.e. less than $2$m/s$^{2}$. It must be also mentioned
that both velocity and brake control inputs are normalized to the interval
$[-1,1]$, where positive values mean throttle actions and the negative, brake ones.

\subsection{Throttle Control}

In previous works, some traditional PI controllers have been designed (refer
e.g. to \cite{Villagra10}), and in \cite{Ines_2011} a fractional order PI
controller was proposed. A fractional order PI controller can be represented
as follows:
\begin{center}%
\begin{equation}
C(s)=k_{p_{1}}+\frac{k_{i}}{s^{\alpha}}=k_{p_{1}}\left(  1+\frac{z_{c}%
}{s^{\alpha}}\right)  ,\text{ with }z_{c}=k_{i}/k_{p_{1}}. \label{PIC}%
\end{equation}

\end{center}

Let assume that the gain crossover frequency is given by $\omega_{c}$, the
phase margin is specified by $\varphi_{m}$ and the output disturbance
rejection is defined by a desired value of a sensitivity function $S(s)$ for a
desired frequencies range. For meeting the system stability and robustness,
the three specifications to fulfill are the following:
\begin{description}
\item[1.] Phase margin specification:
\begin{equation}
Arg[G_{ol}(j\omega_{cp})]=Arg[C(j\omega_{cp})G(j\omega_{cp})]=-\pi+\varphi
_{m}. \label{s1}%
\end{equation}
\item[2.] Gain crossover frequency specification:
\begin{equation}
\left\vert G_{ol}(j\omega_{cp})\right\vert =\left\vert C(j\omega_{cp})G(j\omega_{cp})\right\vert =1. \label{s22}%
\end{equation}
\item[3.] Output disturbance rejection for $\omega\leq\omega_{s}=0.035$rad/s:
\begin{equation}
\left\vert S(j\omega)\right\vert _{dB}=\left\vert \frac{1}{1+C(j\omega
)G(j\omega)}\right\vert _{dB}\leq-20dB,\text{ }\omega\leq\omega_{s}.
\label{s3}%
\end{equation}
\end{description}

Using these three specification and solving the following equations the
controller parameters will be obtained \cite{Ines_2011}.

\begin{center}%
\begin{equation}
z_{c}=\frac{-\tan\left[  \arctan\left(  \frac{\omega_{cp}}{p}\right)
+\varphi_{m}\right]  }{\omega_{cp}^{-\alpha}\left\{  \sin\phi+\cos\phi
\tan\left[  \arctan\left(  \frac{\omega_{cp}}{p}\right)  +\varphi_{m}\right]
\right\}  }. \label{spe1}%
\end{equation}
\[
\frac{Kk_{p_{1}}\sqrt{\left(  1+z_{c}\omega_{cp}^{-\alpha}\cos\phi\right)
^{2}+\left(  z_{c}\omega_{c}^{-\alpha}\sin\phi\right)  ^{2}}}{\sqrt
{\omega_{cp}^{2}+p_{{}}^{2}}}=1
\]
\begin{equation}
k_{p_{1}}^{2}+k_{i}^{2}\omega_{cp}^{-2\alpha}+2k_{p_{1}}k_{i}\omega
_{cp}^{-\alpha}\cos\phi=\frac{\omega_{cp}^{2}+p_{{}}^{2}}{K_{{}}^{2}}.
\label{spe2}%
\end{equation}
\begin{equation}
\left\vert S\right\vert =\left\vert \frac{1}{1+k_{p_{1}}\left[  1+z_{c}%
\omega^{-\alpha}\cos\phi-jz_{c}\omega^{-\alpha}\sin\phi\right]  \left(
\frac{K}{j\omega+p}\right)  }\right\vert . \label{spe3}%
\end{equation}
\end{center}

Solving the set of equations (\ref{spe1}), (\ref{spe2}) and (\ref{spe3}) with
the Matlab function \textit{fsolve}, the values of the controller parameters
are: $k_{p1}=0.09$, $k_{i}=0.025$ and $\alpha=0.8$.

\subsection{Brake Control}

As shown, when the brake is active, vehicle dynamics is different, so the
brake pedal needs its own controller. In particular, in order to have a robust
controller for the uncertain model identified in (\ref{brake_eq}), a robust
fractional order PI controller is designed based on the method proposed in
\cite{Monje_2003}. For the purpose of robustness to time constant variations,
the gain and phase margins have been also taken as the main indicators. Thus,
the specifications to meet are the ones in (\ref{s1}) and (\ref{s22}),
referred to phase margin ($\phi_{m}$) and phase crossover frequency
($\omega_{cp}$) specifications, and the one referring to gain margin ($M_{g}%
$), i.e.:
\begin{align}
Arg\left(  C(j\omega_{cg})G(j\omega_{cg})\right)   &  =-\pi,\label{Sp3}\\
\left\vert C(j\omega_{cg})G(j\omega_{cg})\right\vert _{dB}  &  =1/M_{g},
\label{Sp4}%
\end{align}
where $\omega_{cg}$ is the gain crossover frequency. Replacing (\ref{brake_eq}%
) and (\ref{PIC}) in equations (\ref{s1}), (\ref{s22}), (\ref{Sp3}) and
(\ref{Sp4}), brake specifications can be given by the following set of four
nonlinear equations with the four unknown variables ($k_{p_{1}},k_{i}%
,\alpha,\omega_{cg}$): {\small
\begin{align}
\arctan\left(  \frac{k_{p_{1}}\omega_{cp}^{\alpha}\sin\frac{\alpha\pi}{2}%
}{k_{i}+k_{p_{1}}\omega_{cp}^{\alpha}\cos\frac{\alpha\pi}{2}}\right)
-\arctan\left(  \tau\omega_{cp}\right)  +\frac{(2-\alpha)\pi}{2}-\phi_{m}  &
=0,\label{N1}\\
\arctan\left(  \frac{k_{p_{1}}\omega_{cg}^{\alpha}\sin\frac{\alpha\pi}{2}%
}{k_{i}+k_{p_{1}}\omega_{cg}^{\alpha}\cos\frac{\alpha\pi}{2}}\right)
-\arctan\left(  \tau\omega_{cg}\right)  +\frac{(2-\alpha)\pi}{2}  &
=0,\label{N2}\\
20\log\left(  \frac{\sqrt{(k_{i}+k_{p_{1}}\omega_{cp}^{\alpha}\cos\frac
{\alpha\pi}{2})^{2}+(k_{p_{1}}\omega_{cp}^{\alpha}\sin\frac{\alpha\pi}{2}%
)^{2}}}{\omega_{cp}^{\alpha}\sqrt{(\tau\omega_{cp})^{2}+1}}\right)   &
=0,\label{N3}\\
20\log\left(  \frac{\sqrt{(k_{i}+k_{p_{1}}\omega_{cg}^{\alpha}\cos\frac
{\alpha\pi}{2})^{2}+(k_{p_{1}}\omega_{cg}^{\alpha}\sin\frac{\alpha\pi}{2}%
)^{2}}}{\omega_{cg}^{\alpha}\sqrt{(\tau\omega_{cg})^{2}+1}}\right)  -\frac
{1}{M_{g}}  &  =0. \label{N4}%
\end{align}
}

{\normalsize To reach out its solution, the Matlab function \textit{FMINCON}
was used, which finds the constrained minimum of a function of several
variables. In this case, (\ref{N3}) was considered as the main function whose
parameters are optimized taking into account (\ref{N1}), (\ref{N2}) and
(\ref{N4}) as its constraints and setting $\phi_{m},$ $\omega_{cp}$ and
$M_{g}$ to $90$deg, $0.3$rad/s and $4$, respectively. Thus, the obtained
controller parameters are: $k_{p_{1}}=0.07$, $k_{i}=0.11$ and $\alpha=0.45$. }

{\normalsize Fig. \ref{Bode_FOPI} shows the Bode plots of the controlled
system by applying the designed controller. As it can be observed, the cross
over frequency is $\omega_{cp}=$ $0.7$rad/s and the phase margin is $\phi
_{m}=93$deg, fulfilling the design specifications. Moreover, the system is
robust to the time constant variation, which is also fulfilled as illustrated
in Fig. \ref{Bode_comp}. }
\begin{center}
{\normalsize \begin{figure}[ptbh]
\begin{center}
{\normalsize \includegraphics[width=0.5\textwidth]{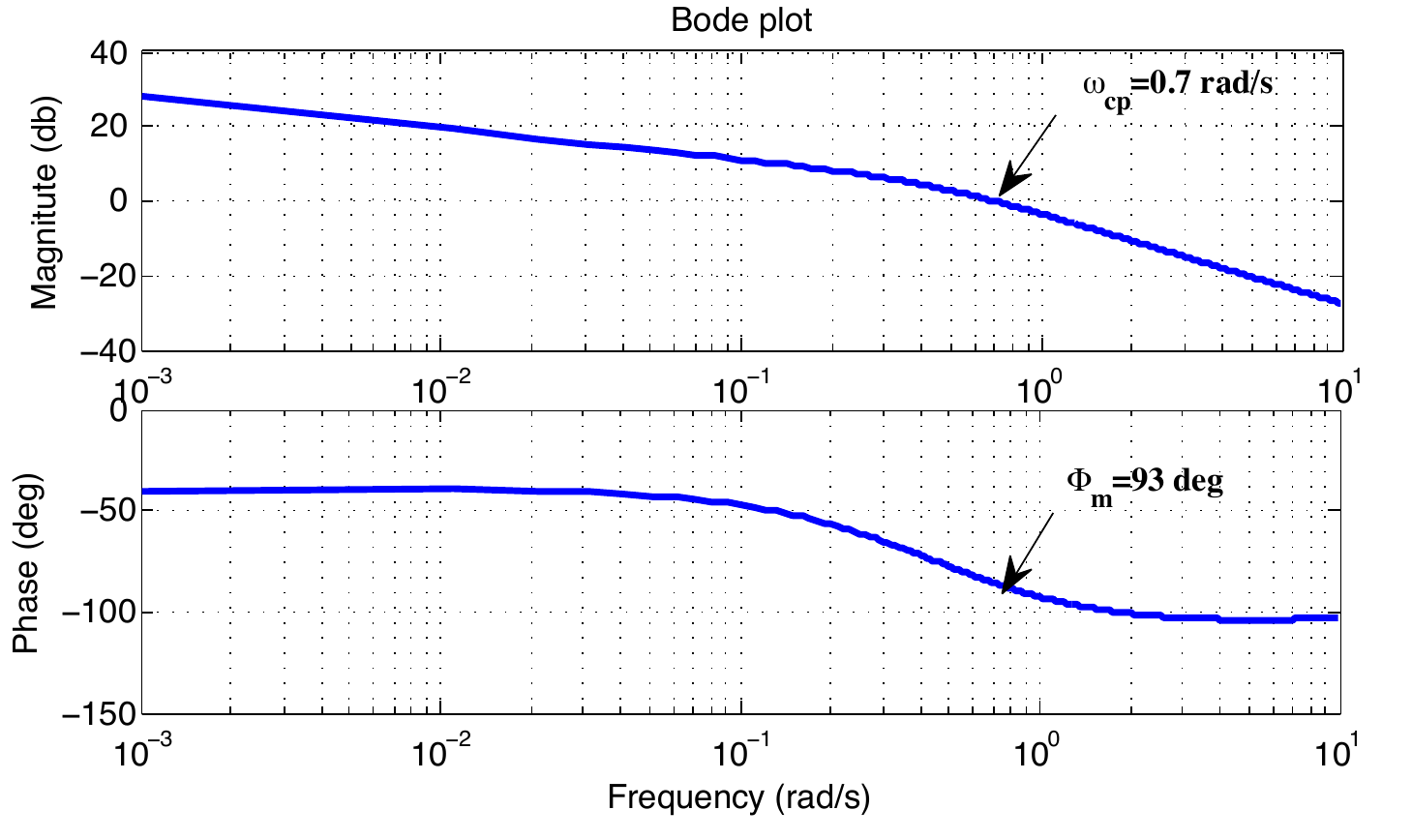} }
\end{center}
\par
{\normalsize  }\caption{Bode plots of the controlled vehicle by applying the
designed PI$^{\alpha}$ brake controller}%
\label{Bode_FOPI}%
\end{figure}\begin{figure}[ptbh]
\begin{center}
{\normalsize \includegraphics[width=0.5\textwidth]{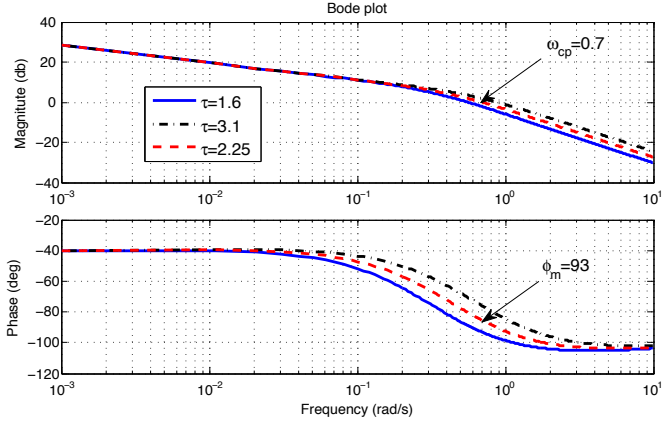} }
\end{center}
\par
{\normalsize  }\caption{Comparison of Bode plots of the controlled vehicle by
applying the designed PI$^{\alpha}$ brake controller with different values of
$\tau$}%
\label{Bode_comp}%
\end{figure}}
\end{center}
{\normalsize Table \ref{cont_PI} summarizes the parameters of the controllers
for brake and throttle control. \begin{table}[pth]
\caption{Parameters of the PI$^{\alpha}$ controller for CC}%
\label{cont_PI}
\begin{center}
{\normalsize
\begin{tabular}
[c]{|l|c|c|c|}\hline
Controller for & $k_{p1}$ & $k_{i}$ & $\alpha$\\\hline
Throttle & 0.09 & $0.025 $ & $0.8$\\\hline
Brake & 0.07 & $0.11 $ & $0.45$\\\hline
\end{tabular}
}
\end{center}
\end{table}}

\subsection{{\protect\normalsize Hybrid Modeling of the Controlled System}}

{\normalsize As mentioned before, velocity are controlled with fractional
order PI in both brake and throttle actions. Regarding to throttle and brake
dynamics, consider a first order system with two different dynamics as
follows:
\begin{equation}
G_{i}(s)=\frac{K_{i}}{s+\tau_{i}}, i=1, 2,
\end{equation}
}

{\normalsize Now consider following fractional order PI to control this
system,
\begin{equation}
C_{i}(s)=k_{p_{i}}+\frac{k_{i_{i}}}{s^{\alpha_{i}}}, i=1, 2.
\end{equation}
where the the parameters regarding to throttle and brake dynamics are shown in following table:  \begin{table}[pth]
\caption{Parameters of the system and controller in throttle and brake action}%
\label{Param_sys}
\begin{center}
{\normalsize
\begin{tabular}
[c]{|l|c|c|c|c|c|}\hline
& $k_{p_{1_{i}}}$ & $k_{i_{I}}$ & $\alpha_{i}$ & $K_{i}$ & $\tau_{i}$\\\hline
Throttle $(i=1)$ & 0.09 & $0.025 $ & $0.8$ & $4.39$ & $0.1746$\\\hline
Brake $(i=2)$ & 0.07 & $0.11 $ & $0.45$ & $\frac{1}{\tau}$ & $\frac{1}{\tau}%
$\\\hline
\end{tabular}
}
\end{center}
\end{table}\newline where $\tau\in\lbrack1.6,3.1]$, concerning to the brake
action. Closed loop transfer function of the system can be represent as,
\begin{equation}
\frac{Y(s)}{R(s)}=\frac{a_{i}s^{\alpha_{i}}+b_{i}}{s^{\alpha_{i}+1}+(\tau
_{i}+a_{i})s^{\alpha_{i}}+b_{i}}, i=1, 2.
\label{TFCL}
\end{equation}
where $a_{i}=K_{i}k_{p_{1_{i}}}$ and $b_{i}=K_{i}k_{i_{i}}$. Assuming
$\alpha_{i}=\frac{p_{i}}{q_{i}}$, (\ref{TFCL}) can be represented as state space:
}{
\begin{align}
D^{\frac{1}{q_{i}}}{x}=\mathcal{A}_i {x}+\mathcal{B}_i r(t), \nonumber\\%
y=\mathcal{C}_i {x}.
\end{align}
}{\normalsize 
where ${x}=\begin{bmatrix} x_{1} & x_{2}& \cdots &  x_{p_{i}+1} & \cdots & x_{p_{i}+q_{i}-1} & x_{p_{i}+q_{i}} \end{bmatrix}^T$ , $\mathcal{A}_i=\begin{bmatrix}
0 & 1 & 0 & \cdots & 0 & \cdots & 0\\
0 & 0 & 1 & \cdots & 0 & \cdots & 0\\
\vdots & \vdots & \vdots & \ddots & \vdots & \ddots & \vdots\\
0 & 0 & 0 & \cdots & 1 & \cdots & 0\\
\vdots & \vdots & \vdots & \ddots & \vdots & \ddots & \vdots\\
0 & 0 & 0 & \cdots & 0 & \cdots & 1\\
-b_{i} & 0 & 0 & \cdots & -(\tau_{i}+a_{i}) & \cdots & 0
\end{bmatrix}$, $\mathcal{B}_i=\begin{bmatrix}
0\\
0\\
0\\
\vdots\\
0\\
\vdots\\
1
\end{bmatrix}$ and $\mathcal{C}_i=\begin{bmatrix} b_i & 0 & \cdots&  a_i & \cdots & 0 & 0 \end{bmatrix}$. 
Now assume that the controller one i.e. $c_{1}(e)$ will be activated if
$e=r(t)-y(t)>-\varepsilon$, ($r(t)=V_{ref}(t)$ and $y(t)=V(t)$) and the other
controller i.e. $c_{2}(e)$ will be activated if $e=r(t)-y(t)<\varepsilon$.
Thus, the flow set and the flow map are taken to be,
\begin{equation}%
D^{\frac{1}{q_{i}}} \begin{bmatrix}
 x\\
 i
\end{bmatrix}
=%
\begin{bmatrix}
\mathcal{A}_{i}x+\mathcal{B}_{i}r_{i}(t)\\
0
\end{bmatrix},
\end{equation}
\begin{align}
C:= \{ (x,i) \in\mathbb{R}^{\alpha_{i}+1} \times\left\{ 1, 2 \right\}  | i=1
\text{ \& }\\
y(t)<r(t)+\varepsilon\text{ or } i=2 \text{ \& } y(t)>r(t)-\varepsilon
\}.\nonumber
\end{align}}

{\normalsize The jump set is taken to be:
\begin{align}
D:= \{ (x,i) \in\mathbb{R}^{\alpha_{i}+1} \times\left\{ 1, 2 \right\}  | i=1
\text{ \& }\\
y(t)=r(t)+\varepsilon\text{ or } i=2 \text{ \& } y(t)=r(t)-\varepsilon
\}.\nonumber
\end{align}
}

{\normalsize Regarding the jump map, since the role of jump changes is to
toggle the logic mode and since the state component $x$ does not change during
jumps, the jump map will be
\begin{equation}%
\begin{bmatrix}
x\\
i
\end{bmatrix}
^{+} =%
\begin{bmatrix}
x\\
3-i
\end{bmatrix}
.
\end{equation}
}

{\normalsize Fig. \ref{Switching} shows the switching between throttle and
brake action corresponding to the $\varepsilon=0$. It is obvious that the
systems is stable during switching of throttle or brake action. $S_{1}$
represent the region where the throttle is active and $S_{2}$ shows the region
where the brake is active. }

{\normalsize \begin{figure}[ptbh]
\begin{center}
{\normalsize \includegraphics[width=0.45\textwidth]{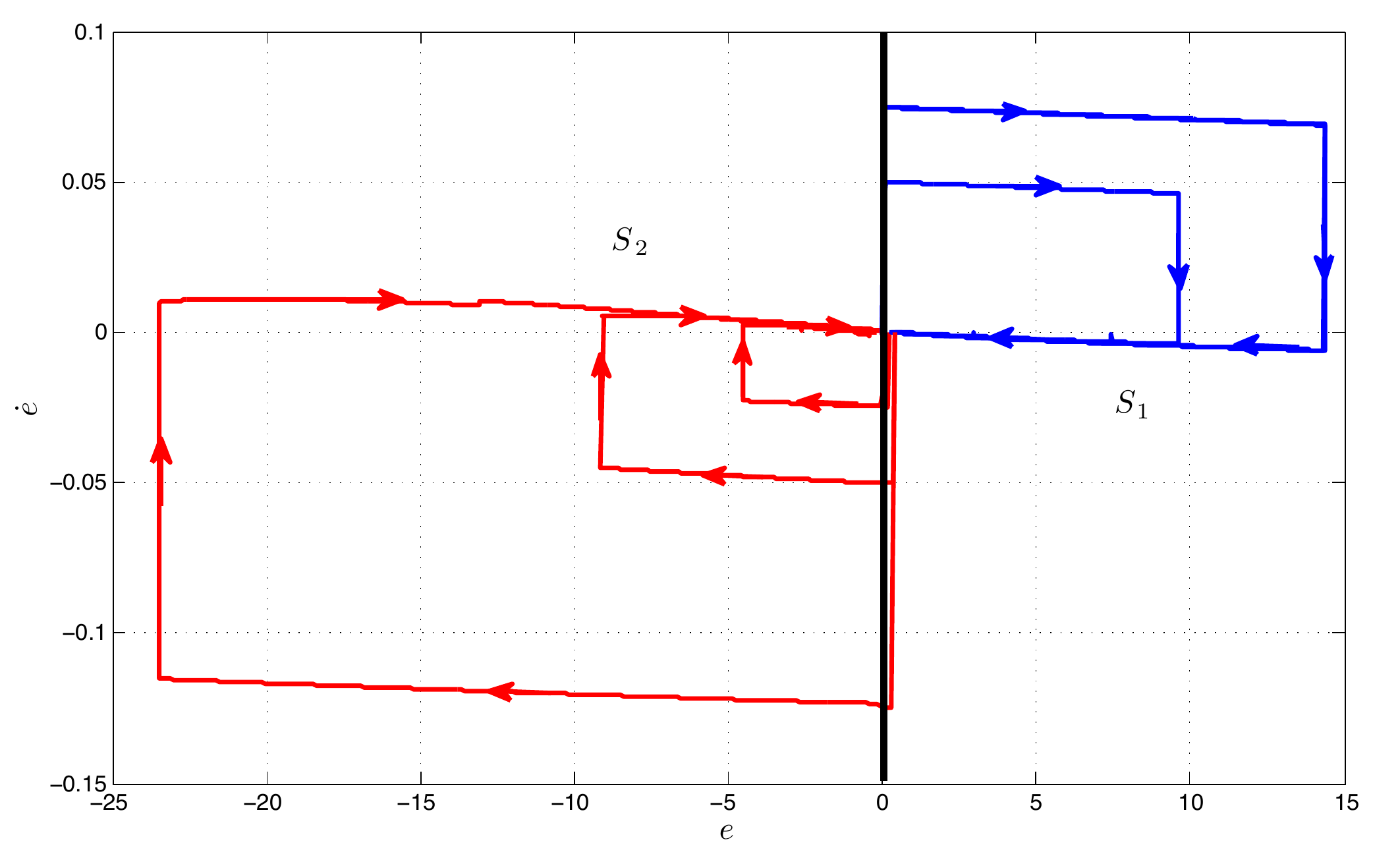} }
\end{center}
\caption{Switching phases}%
\label{Switching}%
\end{figure}}

\section{Experimental and Simulation Results\label{results}}

\begin{figure}[ptbh]
\begin{center}
{\normalsize \includegraphics[width=0.4\textwidth]{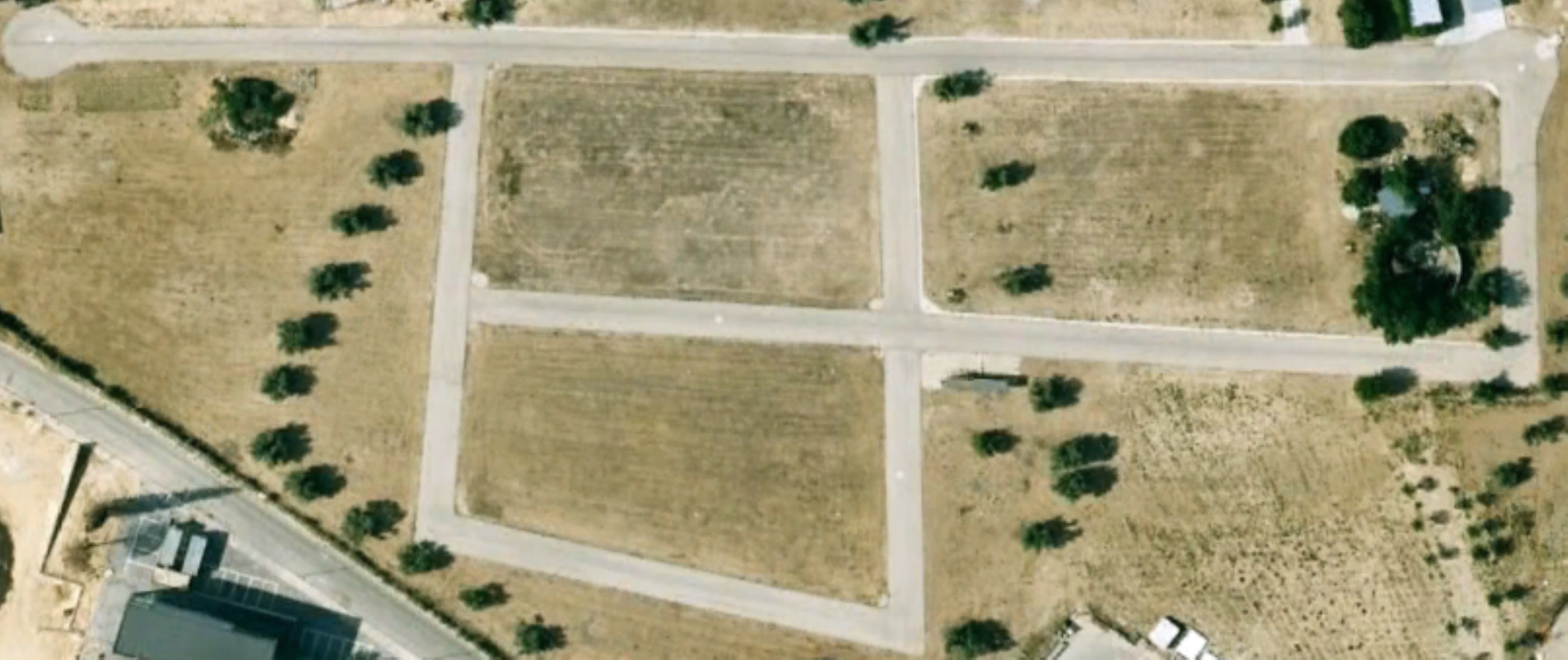}  }
\end{center}
\par
{\normalsize  }\caption{Private driving circuit at the Center for Automation
and Robotics}%
\label{circuit}%
\end{figure}

The designed controllers have been tested in simulation and on the real
vehicle in the CAR's private driving circuit illustrated in Fig.
\ref{circuit}. This circuit has been designed with scientific purposes so only
experimental vehicles are driven in this area. It includes $90\deg$ bends and
different slopes so as to validate the controller in different circumstances
as close to a real environment as possible. 

{\normalsize The simulation results are carried out in the MATLAB/Simulink
environment. In order to show the efficiency of the robust controller, a
random noise with mean value of }${\normalsize 0.85}${\normalsize  is added to
nominal value of $\tau=2.25$. To show the brake action and the performance of
the robust controller which is designed for uncertain system (\ref{brake_eq}),
the car moved with fixed pedal to reach the velocity of }${\normalsize 30}%
${\normalsize km/h; then, the brake controller is activated. The experimental
brake results are shown in Fig. \ref{BCS}. As aforementioned, the main
important limitation which have been considered is acceleration i.e. $[-2,2]$.
Regarding to this results it is obvious that the application of the robust
fractional order controller fulfills the acceleration limitation for the
system. \begin{figure}[ptbh]
\begin{center}
{\normalsize \includegraphics [width=0.5\textwidth]{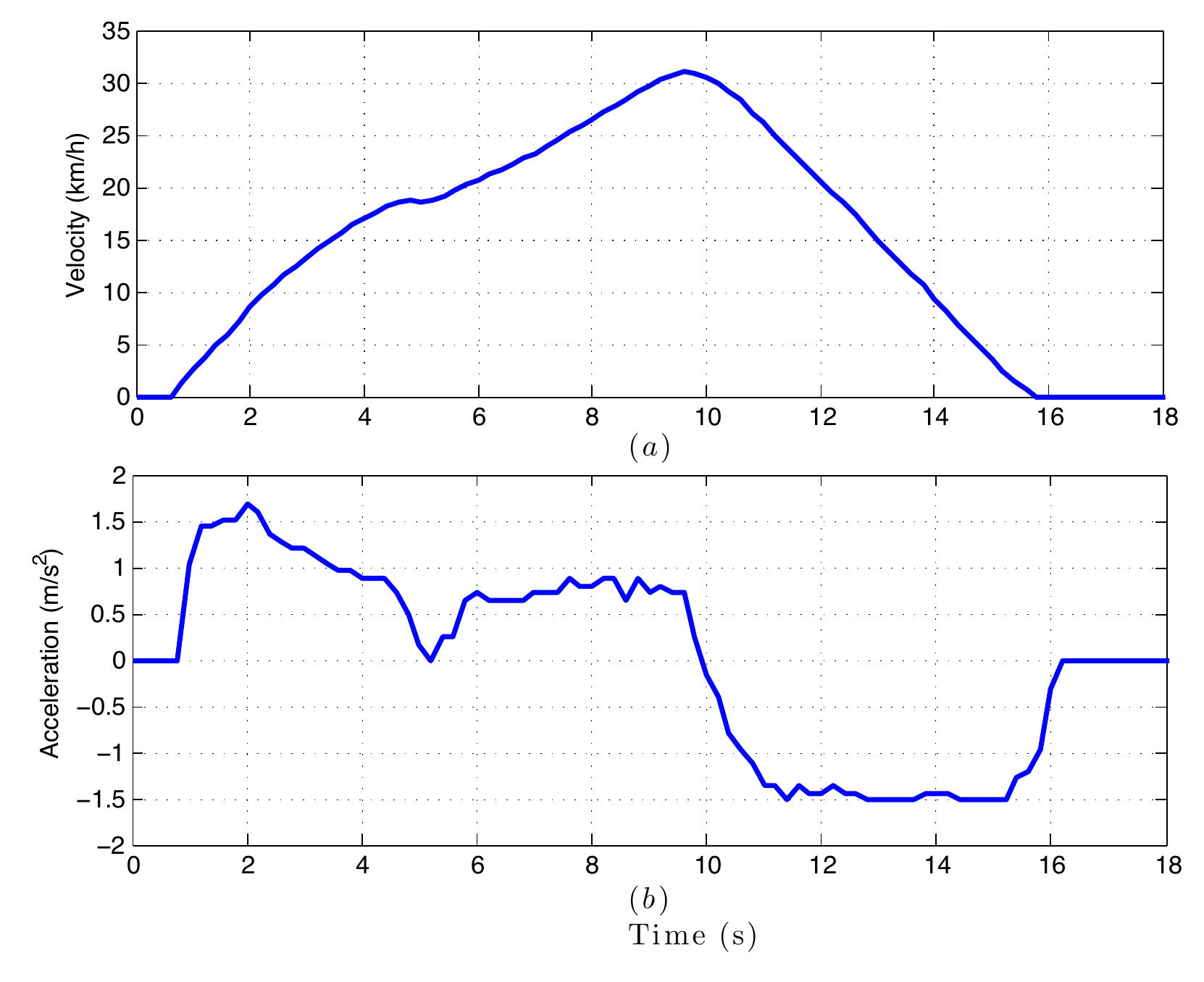}  }
\end{center}
\par
{\normalsize  }\caption{Brake Control: $(a)$ velocity, $(b)$ acceleration}%
\label{BCS}%
\end{figure}}

{\normalsize Fig. \ref{TB1} (Experiment I) and \ref{TB2} (Experiment II) show
a comparison between simulation and experimental results of the controlled
vehicle for two different variable speed references. Velocity tracking,
acceleration and normalized control action are shown in each figures. As it
can be seen, the results show that the acceleration and control action are met
the desired intervals. One can appreciate the soft action over vehicle's
actuators obtaining a good comfort for car's occupants --this is reflected in
the acceleration values. Moreover, the experimental results are tracked the
reference value in both brake and throttle actions which is also verify the
simulations results. }

{\normalsize To sum up, fractional order hybrid controllers can be useful
controllers to control autonomous vehicles in the brake and throttle action,
specially due to its possibility of obtaining more adjustable time and
frequency responses and allowing the fulfillment of more robust performances.
The vehicle behavior is significantly good, especially concerning the comfort
of vehicle's occupants. }

{\normalsize \begin{figure}[ptbh]
\begin{center}
{\normalsize \includegraphics[width=0.5\textwidth]{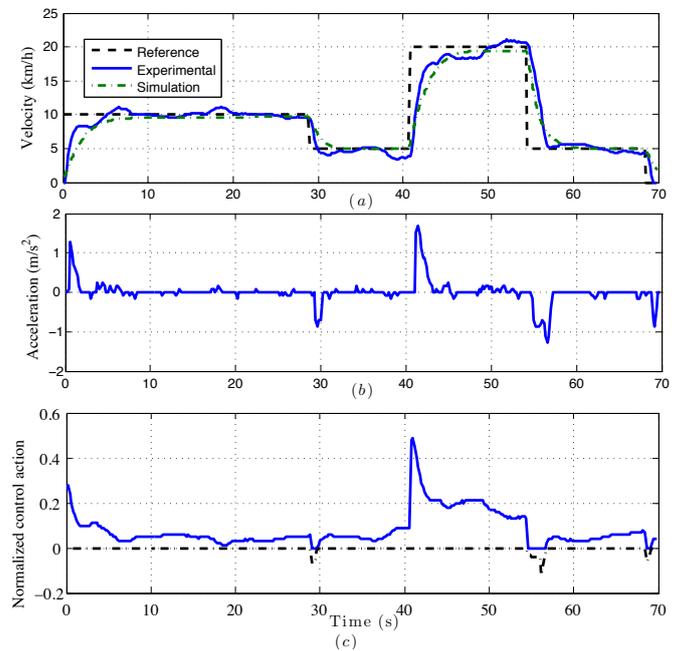}  }
\end{center}
\par
{\normalsize  }\caption{Cruise control results for Experiment I: $(a)$
velocity, $(b)$ acceleration, $(c)$ normalized control action}%
\label{TB1}%
\end{figure}}

{\normalsize \begin{figure}[ptbh]
\begin{center}
{\normalsize \includegraphics[width=0.5\textwidth]{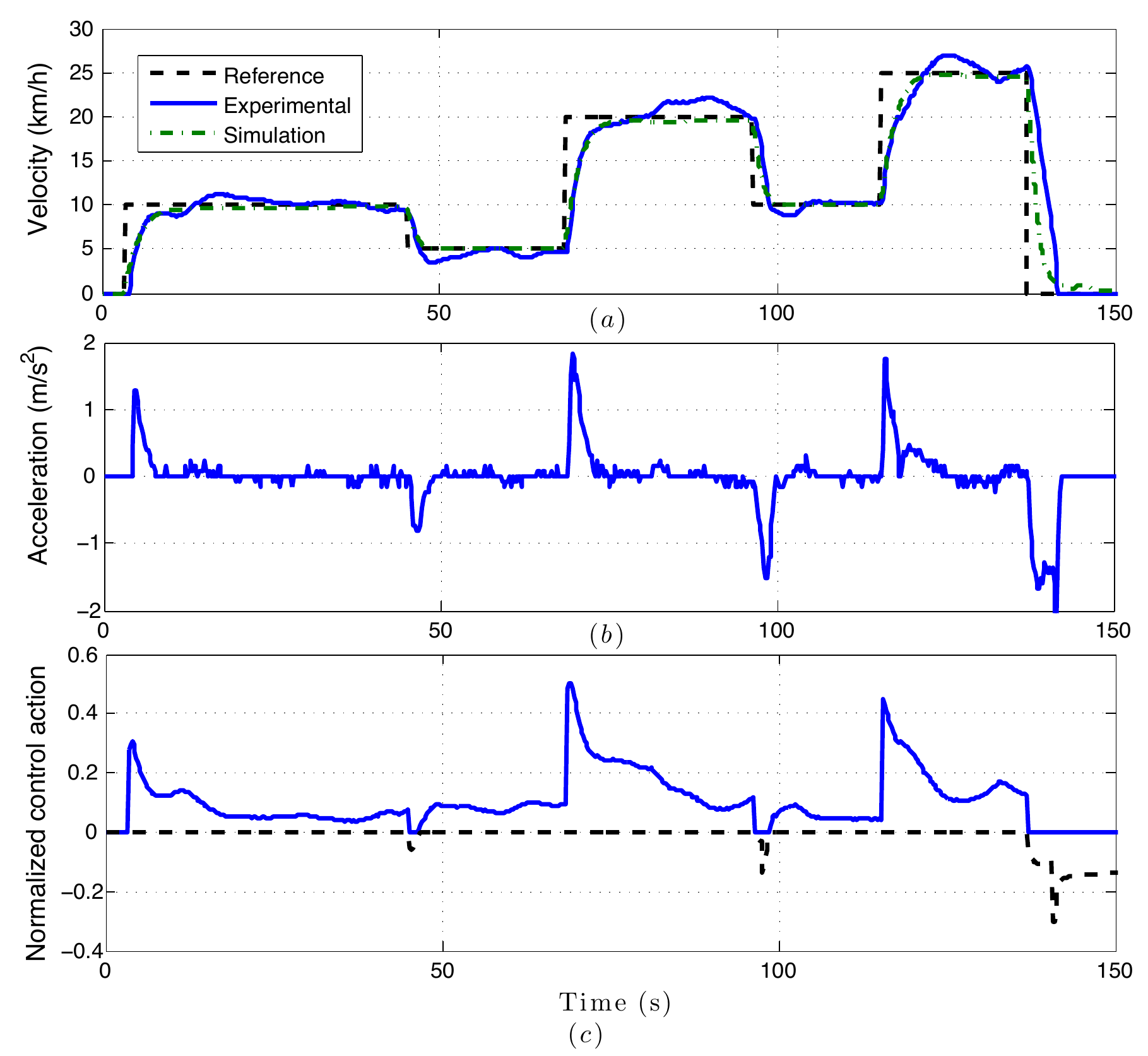}  }
\end{center}
\par
{\normalsize  }\caption{Cruise control results for Experiment II: $(a)$
velocity, $(b)$ acceleration, $(c)$ normalized control action}%
\label{TB2}%
\end{figure}}

\subsection{Digital implementation of fractional order controllers}

It has to be taken into account that a fractional order controller is an
infinite-dimensional linear filter, and that all existing implementation
schemes are based on finite-dimensional approximations. In practice, we use a
digital method, specifically the indirect discretization method, which
requires two steps: firstly obtaining a finite-dimensional continuous
approximation, and secondly discretizing the resulting $s$-transfer function.
In our case, in order to preserve the integral effect, the integral part
$s^{-\alpha}$ has been implemented as follows: $s^{-\alpha}=s^{-1}s^{1-\alpha
}.$Therefore, only the fractional part $R_{d}(s)=s^{1-\alpha}$ has been
approximated. 

To obtain a finite-dimensional continuous approximation of the fractional
order differentiator, the modified Oustaloup's method is used (see e.g.
\cite{Monje10}). Thus, an integer order transfer function that fits the
frequency response of $R_{d}(s)$ in the range $\omega\in(10^{-3},10^{3})$ is
obtained with 7 poles and 7 zeros. Later, the discretization of this
continuous approximation is carried out by using the Tustin rule with a
sampling period $T_{s}=0.2$s -GPS sampling period-, obtaining the following
digital IIR filters: 

\begin{itemize}
\item Throttle controller:%
\begin{equation}
R_{dT}(z)=\frac{\underset{k=0}{\overset{7}{\sum}}b_{k}z^{-k}}{1+\underset
{k=1}{\overset{7}{\sum}}a_{k}z^{-k}},\label{filter}%
\end{equation}
where $b_{0}=0.1573,$ $b_{1}=0.1325$, $b_{2}=-0.4389$, $b_{3}=-0.3658$,
$b_{4}=0.406$, $b_{5}=0.3342$, $b_{6}=-0.1244$, $b_{7}=-0.1009$,
$a_{1}=-0.8662$, $a_{2}=-2.746$, $a_{3}=2.339$, $a_{4}=2.507$, $a_{5}=-2.095$,
$a_{6}=-0.7602$ and $a_{7}=0.6211$. Therefore, the resulting total fractional
order controller is an 8th-order digital IIR filter given by:%
\[
C_{T}(z)=0.09+0.025\left(  \frac{1-z^{-1}}{T_{s}}\right)  R_{dT}(z).
\]
\item Brake controller.- The fractional part of this controller $R_{dB}(z)$
was implemented by ({\normalsize \ref{filter}}) but with the following
coefficients: $b_{0}=0.3529,$ $b_{1}=0.1878$, $b_{2}=-1.0274$, $b_{3}%
=-0.5381$, $b_{4}=0.9959$, $b_{5}=0.5128$, $b_{6}=-0.3215$, $b_{7}=-0.1625$,
$a_{1}=-0.5400$, $a_{2}=-2.88062$, $a_{3}=1.5053$, $a_{4}=2.7658$,
$a_{5}=-1.3952$, $a_{6}=-0.8852$ and $a_{7}=0.4299$. In this case, the
resulting total fractional order controller is an 8th-order digital IIR filter
given by:%
\[
C_{B}(z)=0.07+0.11\left(  \frac{1-z^{-1}}{T_{s}}\right)  R_{dB}(z).
\]
\end{itemize}

\section{{\protect\normalsize Conclusion\label{conclu}}}

{\normalsize A hybrid fractional order controller is proposed to control the velocity of the car in low speed. The system is modeled as a hybrid differential inclusions. The system has different dynamics during the acceleration and deceleration of the car, and therefore different controllers are needed. A fractional order PI is used to control the throttle action, whereas a robust fractional order PI controller is designed for the uncertain model identified regarding to different brake actions. Both simulated and experimental results show the
efficiency of the proposed strategy. }

\end{document}